\documentclass[showpacs,amsmath,amssymb,twocolumn,prl,superscriptaddress,longbibliography]{revtex4-1}

\usepackage{qcircuit}
\usepackage[dvips]{graphicx}
\usepackage{amsmath,amssymb,amsthm,mathrsfs,amsfonts,dsfont}
\usepackage{subfigure, epsfig}
\usepackage{braket}
\usepackage{bm}
\usepackage{enumerate}
\usepackage{color}
\usepackage{placeins}

\newcommand{\tr}{\mathrm{Tr}}

\usepackage{soul}
\newcommand{\newedit}[1]{{#1}}

\usepackage{ulem}

\begin{document}

\title{Variational {ansatz-based} quantum simulation of imaginary time evolution}
\begin{abstract}
Imaginary time evolution is a powerful tool for studying quantum systems. While it is \newedit{possible} to simulate with a classical computer, the time and memory requirements generally scale exponentially with the system size. Conversely, quantum computers can efficiently simulate quantum systems, but not non-unitary imaginary time evolution. We propose a variational algorithm for simulating imaginary time evolution on a hybrid quantum computer. We use this algorithm to find the ground state energy of many-particle systems; specifically molecular Hydrogen and Lithium Hydride, finding the ground state with high probability. Our method can also be applied to general optimisation problems and quantum machine learning. As our algorithm is hybrid, suitable for error mitigation, and can exploit shallow quantum circuits, it can be implemented with current quantum computers. 
\end{abstract}

\date{\today}

\author{Sam McArdle}
\affiliation{Department of Materials, University of Oxford, Parks Road, Oxford OX1 3PH, United Kingdom}

\author{Tyson Jones}
\affiliation{Department of Materials, University of Oxford, Parks Road, Oxford OX1 3PH, United Kingdom}

\author{Suguru Endo}
\affiliation{Department of Materials, University of Oxford, Parks Road, Oxford OX1 3PH, United Kingdom}

\author{Ying Li}
\affiliation{Graduate School of China Academy of Engineering Physics, Beijing 100193, China}

\author{Simon Benjamin}
\affiliation{Department of Materials, University of Oxford, Parks Road, Oxford OX1 3PH, United Kingdom}

\author{Xiao Yuan}
\email{xiao.yuan.ph@gmail.com}
\affiliation{Department of Materials, University of Oxford, Parks Road, Oxford OX1 3PH, United Kingdom}

\maketitle



\noindent\textbf{INTRODUCTION}\\
Imaginary time is an unphysical, yet powerful, mathematical concept. 
It has been utilised in numerous physical domains including: quantum mechanics, statistical mechanics, and cosmology. 
Often referred to as performing a `Wick rotation' \cite{PhysRev.96.1124}, replacing real time with imaginary time connects Euclidean and Minkowski space~\cite{Poincara1906}, quantum and statistical mechanics~\cite{sakurai2017modern}, and static problems to problems of dynamics~\cite{e17020772}.
In quantum mechanics, propagating a wavefunction in imaginary time enables: the study of finite temperature properties~\cite{Verstraete04,Zwolak04,PhysRevX.5.041032}, finding the ground state wavefunction and energy {(such as in density matrix renormalisation group)}~\cite{LEHTOVAARA2007148,Christina10,C5RA23047K,SHI2018245}, and simulating real time dynamics {(such as time dependent Hartree)}~\cite{McCleanE3901,McCleanReal15}. For a system with Hamiltonian, $H$, evolving in real time, $t$, the propagator is given by $e^{-iHt}$. The corresponding propagator in imaginary time, $\tau = it$, is given by $e^{-H\tau}$; a non-unitary operator. 

Using a classical computer, we can simulate imaginary time evolution by evaluating the propagator and applying it to the system wavefunction. {There also exist various related classical methods, such as quantum Monte Carlo \cite{al2006auxiliary, motta2018ab} and density matrix renormalization group \cite{chan2011density, stoudenmire2017sliced} for solving different problems. } However, because the dimension of the wavefunction grows exponentially with the number of particles, classical simulation of many-body quantum systems is generally hard~\cite{Feynman1982}. While efficient variational trial states have been developed for a number of applications~\cite{RevModPhys.71.463}, powerful trial wavefunctions typically require classical computational resources which scale exponentially with the system size~\cite{SHI2018245}.

Quantum computing can naturally and efficiently store many-body quantum states, and hence is suitable for simulating quantum systems \cite{RevModPhys.86.153}. We can map the system Hamiltonian to a qubit Hamiltonian, and simulate real time evolution (as described by the Schr\"odinger equation) by realising the corresponding unitary evolution with a quantum circuit~\cite{nielsen2002quantum}. Using Trotterization~\cite{trotter1959product}, the real time propagator can be decomposed into a sequence of single and two qubit gates~ \cite{Abrams97}. The ability to represent the real time propagator with a sequence of gates stems from its unitarity. In contrast, because the imaginary time operator is non-unitary, it is not straightforward to decompose it into a sequence of unitary gates using Trotterization, and thus directly realise it with a quantum circuit. As a result, alternative methods are required to implement imaginary time evolution using a quantum computer.

Classically, we can simulate real (imaginary) time evolution of parametrised trial states by repeatedly solving the (Wick-rotated) Schr\"odinger equation over a small timestep, and updating the parameters for the next timestep~\cite{JACKIW1979158,LEHTOVAARA2007148,Kramer08,Christina10,PhysRevLett.107.070601,SHI2018245,ashida2018variational}. This method has recently been extended to quantum computing, where it was used to simulate real time dynamics~\cite{Li2017}. Closely related are the variational quantum eigensolver (VQE)~\cite{peruzzo2014variational,wang2015quantum,PRXH2,PhysRevA.95.020501,VQETheoryNJP,PhysRevLett.118.100503,kandala2017hardware} and the quantum approximate optimisation algorithm (QAOA)~\cite{farhi2014quantum}, which update the parameters using a classical optimisation routine, to find the minimum energy eigenvalue of a given Hamiltonian. As `hybrid quantum-classical methods', these algorithms use a small quantum computer to carry out a classically intractable subroutine, and a classical computer to solve the higher level problem. The quantum subroutine may only require a small number of qubits and a low depth circuit, presenting a potential use for noisy intermediate-scale quantum hardware~\cite{preskill2018quantum}.

In this paper, we propose a method to simulate imaginary time evolution on a quantum computer, using a hybrid quantum-classical variational algorithm. The proposed method thus combines the power of quantum computers to efficiently represent many-body quantum states, with classical computers' ability to simulate arbitrary (including unphysical) processes. We discuss using this method to find the ground state energy of many-body quantum systems, and to solve optimisation problems. We then numerically test the performance of our algorithm at finding the ground state energy of both the Hydrogen molecule (H$_2$) and Lithium Hydride (LiH). We compare our results for LiH to those obtained using the VQE with gradient descent. As our algorithm only requires a low depth circuit, it can be realised with current and near-term quantum processors.\\

\noindent\textbf{RESULTS}\\
\noindent\textit{Variational imaginary time evolution.}
We focus on many-body systems that are described by Hamiltonians $H = \sum_i \lambda_i h_i$, with real coefficients, $\lambda_i$, and observables, $h_i$, that are tensor products of Pauli matrices. We assume that the number of terms in this Hamiltonian scales polynomially with the system size, which is true for many physical systems, such as molecules or the Fermi-Hubbard model. Given an initial state $\ket{\psi}$, the normalised imaginary time evolution is defined by
\begin{equation}\label{imagine}
\begin{aligned}
		\ket{\psi(\tau)}&=A(\tau){e^{-H\tau}\ket{\psi(0)}},
		\end{aligned}
\end{equation}
where $A(\tau)=1/{\sqrt{\bra{\psi(0)}e^{-2H\tau}\ket{\psi(0)}}}$ is a normalisation factor. In the instance that the initial state is a maximally mixed state, the state at time $\tau$ is a thermal or Gibbs state $\rho_{T = 1/\tau} = e^{-H\tau}/\tr[e^{-H\tau}]$, with temperature $T = 1/\tau$. When the initial state has a non-zero overlap with the ground state, the state at $\tau\rightarrow\infty$ is the ground state of $H$. Equivalently, the Wick rotated Schr\"odinger equation is,
\begin{equation}\label{diffEq}
	\frac{\partial \ket{\psi(\tau)}}{\partial \tau} = -(H - E_\tau)\ket{\psi(\tau)},
\end{equation}
where the term $E_\tau = \braket{{\psi(\tau)}|H|{\psi(\tau)}}$ results from enforcing normalisation. Even if $\ket{\psi(\tau)}$ can be represented by a quantum computer, the non-unitary imaginary time evolution cannot be naively mapped to a quantum circuit. 

In our variational method, instead of directly encoding the quantum state $\ket{\psi(\tau)}$ at time $\tau$, we  approximate it using a parametrised trial state $\ket{\phi(\vec{\theta}(\tau))}$, with $\vec{\theta}(\tau) = (\theta_1(\tau), \theta_2(\tau),\dots, \theta_N(\tau))$. 
This stems from the intuition that the physically relevant states are contained in a small subspace of the full Hilbert space \cite{PhysRevLett.106.170501}.
The trial state is referred to as the ansatz. In condensed matter physics and computational chemistry, a wide variety of ans\"atze have been proposed for both classical and quantum variational methods \cite{verstraete2008matrix,RevModPhys.86.153,SHI2018245,whaley2014quantum}.

Using a quantum circuit, we prepare the trial state, $\ket{\phi(\vec{\theta})}$, by applying a sequence of parametrised unitary gates, $V(\vec{\theta}) = U_{N}(\theta_N)\dots U_{k}(\theta_k)\dots U_{1}(\theta_1)$ to our initial state, $\ket{\bar{0}}$. We express this as $\ket{\phi(\vec{\theta})}= V(\vec{\theta})\ket{\bar{0}}$ and remark that $V(\vec{\theta})$ is also referred to as the ansatz. We refer to all possible states that could be created by the circuit $V$ as the `ansatz space'. Here, $U_{k}(\theta_k)$ is the $k^{\textrm{th}}$ unitary gate, controlled by parameter $\theta_k$, and the gate can be regarded as a single or two qubit gate. 

To simulate the imaginary time evolution of the trial state, we use McLachlan's variational principle~\cite{McLachlan, broeckhove1988equivalence},
\begin{equation}\label{McLachlansEq}
	\begin{aligned}
		\delta \|({\partial}/{\partial \tau} + H-E_\tau)\ket{\psi(\tau)}\|=0,
	\end{aligned}
\end{equation}
where $\|\rho\| = \tr[\sqrt{\rho\rho^\dag}]$ denotes the trace norm of a state. By replacing $\ket{\psi(\tau)}$ with $\ket{\phi(\tau)}=\ket{\phi(\vec{\theta}(\tau))}$, we effectively project the desired imaginary time evolution onto the manifold of the ansatz space. The evolution of the parameters is obtained from the resulting differential equation
\begin{equation}\label{ImagMatrixEq}
	\begin{aligned}
		\sum_jA_{ij}\dot{\theta}_j = C_i,
	\end{aligned}
\end{equation}
where 
\begin{equation}\label{ACdef}
\begin{aligned}
	A_{ij} &= \Re\left(\frac{\partial \bra{\phi(\tau)}}{\partial \theta_i}\frac{\partial \ket{\phi(\tau)}}{\partial \theta_j}\right),\\
	C_i &=\Re\left( -\sum_\alpha \lambda_\alpha \frac{\partial \bra{\phi(\tau)}}{\partial \theta_i} h_\alpha \ket{\phi(\tau)}\right),
\end{aligned}
\end{equation}
and $h_\alpha$ and $\lambda_\alpha$ are the Pauli terms and coefficients of the Hamiltonian, as described above. The derivation of Eq.~\eqref{ImagMatrixEq} can be found in the Supplementary Materials. As both $A_{ij}$ and $C_i$ are real, the derivative $\dot{\theta}_j$ is also real, as required for parametrising a quantum circuit. Interestingly, although the average energy term $E_\tau$ appears in Eq.~\eqref{diffEq}, it does not appear in Eq.~\eqref{ImagMatrixEq}. This is because the ansatz applied maintains normalisation, as it is composed of unitary operators. \\

\noindent\textit{Imaginary time evolution with quantum circuits.}
By following a similar method to that introduced in Ref.~\cite{Li2017}, we can efficiently measure $A_{ij}$ and $C_i$ using a quantum computer. We assume that the derivative of a unitary gate $U_{i}(\theta_i)$ can be expressed as 
	${\partial U_{i}(\theta_i)}/{\partial \theta_i} = \sum_k f_{k,i}U_{i}(\theta_i)\sigma_{k,i}$,
with unitary operator $\sigma_{k,i}$. The derivative of the trial state is given by
	${\partial \ket{\phi(\tau)}}/{\partial \theta_i} = \sum_k f_{k,i}\tilde{V}_{k,i}\ket{\bar{0}}$,
with $\tilde{V}_{k,i} = U_{N}(\theta_N)\dots U_{i+1}(\theta_{i+1})U_{i}(\theta_i)\sigma_{k,i}\dots U_{1}(\theta_1)$. There are typically only one or two terms resulting from each derivative.
As an example, when $U_i(\theta_i)$ is a single qubit rotation $R_z(\theta_i) = e^{-i\theta_i\sigma_z/2}$, the derivative ${\partial U_{i}(\theta_i)}/{\partial \theta_i} = -i/2\times \sigma_z e^{-i\theta_i \sigma_z/2}$. The coefficients $A_{ij}$ and $C_i$ are given by
\begin{equation}
	\begin{aligned}
		A_{ij} &= \Re\left(\sum_{k,l} f_{k,i}^*f_{l,j}\bra{\bar{0}} \tilde{V}_{k,i}^\dag \tilde{V}_{l,j}\ket{\bar{0}}\right),\\
		C_i &= \Re\left(\sum_{k,\alpha} f_{k,i}^*\lambda_\alpha \bra{\bar{0}} \tilde{V}_{k,i}^\dag h_\alpha V\ket{\bar{0}}\right).
			\end{aligned}
\end{equation}
All of these terms are of the form $a{\Re}(e^{i\theta}\bra{\bar{0}}U\ket{\bar{0}})$ and can be evaluated using the circuits shown in the Supplementary Materials.

With $A(\tau)$ and $\vec{C}(\tau)$ at time $\tau$, the imaginary time evolution over a small interval $\delta \tau$ can be simulated by evaluating $\dot{\vec{\theta}}(\tau)=A^{-1}(\tau)\cdot \vec{C}(\tau)$, and using a suitable update rule, such as the Euler method,
\begin{equation}\label{Euler}
	\begin{aligned}
	\vec{\theta}({\tau + \delta \tau}) &\simeq \vec{\theta}(\tau) +  \dot{\vec{\theta}}(\tau)\delta \tau= \vec{\theta}(\tau) + A^{-1}(\tau)\cdot \vec{C}(\tau)\delta \tau.
	\end{aligned}
\end{equation}
By repeating this process $N_T = \tau_{total}/\delta \tau$ times, we can simulate imaginary time evolution over a duration $\tau_{total}$. 
{
Often, the satisfying parameter evolution is not unique and Eq.~\eqref{ImagMatrixEq} is underdetermined. In that case, we can employ truncated singular value decomposition to approximately invert $A$, or Tikhonov regularisation to additionally constrain the parameters to vary smoothly. We elaborate upon these strategies in the Supplementary Materials.
}

{
A limitation of our variational method is that the ansatz may not be able to faithfully describe all states on the desired trajectory, much like its real time counterpart~\cite{Li2017}. Even though such states lie in a small subspace of the full Hilbert space~\cite{PhysRevLett.106.170501}, it is difficult to prove that they can be generated by a given ansatz, despite promising numerical results~\cite{Li2017}. However, our numerical results are similarly promising for imaginary time, and demonstrate it to be a robust routine for energy minimisation. \newedit{Moreover, we believe that when tasked with finding the ground state using imaginary time evolution, a small deviation from the true evolution is less problematic than when trying to simulate real time evolution. This is because imaginary time evolution always drives a state towards the ground state (or one of the lowest eigenstates), whereas the real time evolution of two closely separated states may be very different. Consequently, as long as errors due to an imperfect ansatz do not cause the simulation to become trapped in local minima, we do not mind if the evolution deviates from the path of true imaginary time evolution, as ultimately, it will still be driven towards the ground state.} Nevertheless, designing ans\"atze that are well suited to imaginary time evolution is an interesting open problem.\\
}

\noindent\textit{Ground state energy via imaginary time evolution.}
We apply our method to the problem of finding the ground state energy of a many-body Hamiltonian, $H$. As with the VQE, our goal is to find the values of the parameters, $\vec{\theta}$, which minimise the expectation value of the Hamiltonian
\begin{equation}
	E_{\min} = \min_{\vec{\theta}} \bra{\phi(\vec{\theta})}H\ket{\phi(\vec{\theta})},
\end{equation}
where $\ket{\phi(\vec{\theta})} = V(\vec{\theta}) \ket{\bar{0}}$ is our variational trial state. The VQE solves this problem by using a quantum computer to construct a good ansatz and measure the expectation value of the Hamiltonian, and a classical optimisation routine to obtain new values of the parameters. 
In order to preserve the exponential speedup of the VQE over classical methods, the trial state is constructed using a number of parameters that scales polynomially with the system size. However, because we may need to consider many possible values for each parameter, the total size of the parameter space still scales exponentially with the system size. Moreover, many optimisation algorithms, such as gradient descent, are liable to becoming trapped in local minima. This combination can make the classical optimisation step of the VQE very difficult~\cite{PhysRevA.92.042303}.

As described above, if the initial state has a non-zero overlap with the ground state, true propagation in imaginary time will evolve the system into the ground state, in the limit that $\tau\rightarrow\infty$. Classically, this has been leveraged as a powerful tool to find the ground state energy of quantum systems~\cite{LEHTOVAARA2007148,Christina10,SHI2018245}. Using our method, we can efficiently simulate ansatz-based imaginary time evolution to find the ground state, using a quantum computer. In the numerical simulations described below, we use the Euler method to solve differential equations, which corresponds to the update rule for the parameters shown in Eq.~\eqref{Euler}. We prove in the Supplementary Materials that when $\delta \tau$ is sufficiently small, the average energy of the trial state, $E(\tau) = \bra{\phi(\tau)}H\ket{\phi(\tau)}$, always decreases when following the Euler update rule: $E(\tau+\delta \tau)\le E(\tau)$.

In this work, we consider gradient descent, a canonical classical optimisation method
\begin{equation}\label{updateGradient}
\begin{aligned}
\vec{\theta}({\tau+\delta \tau}) &= \vec{\theta}({\tau}) +  \vec{G}(\tau)\delta \tau = \vec{\theta}({\tau}) +  \vec{C}(\tau)\delta \tau,\\
\end{aligned}	
\end{equation}
where $\vec{G}(\tau) =- \nabla E(\tau)$ is the gradient of $E(\tau)$ and $\vec{C}(\tau)\equiv - \nabla E(\tau)$ is the same vector in Eq.~\eqref{ImagMatrixEq}. Classical optimisation methods only consider information about the average energy, and not about the ansatz itself, which is encoded in the matrix $A$, used only in variational imaginary time evolution. \\

\noindent\textit{Toy example.}
\newedit{
Here we present two simple toy examples which highlight the difference between variational imaginary time evolution and gradient descent for finding the ground state energy of Hamiltonians. Consider the following Hamiltonians}
\begin{equation}
	H_A = \left(
	\begin{array}{cccc}
 	1 &0& 0& 0\\
 	0 &2& 0& 0\\
	0 &0& 3& 0\\
 	0 &0& 0& 0
 	\end{array}\right), \;\;
 	H_B = \left(
	\begin{array}{cccc}
 	1 &0& 0& 0\\
 	0 &1& 0& 0\\
	0 &0& 2& 0\\
 	0 &0& 0& 0
 	\end{array}\right)
\end{equation}
with ans\"atze
\begin{align}
	&\ket{\psi_A(\theta_1, \theta_2, \theta_3)} = e^{i \theta_3} CR_Y^{0,1}(\theta_2) R_X^0(\theta_1) \ket{00}, \\
	&\ket{\psi_B(\theta_1, \theta_2, \theta_3)} = e^{i \theta_3} CR_Y^{0,1}(\theta_2) R_X^0(\theta_1)  R_X^1(\theta_1) \ket{01},
\end{align}
prepared by circuits ((A) and (B) respectively)
\begin{align*}
&\Qcircuit @C=.7em @R=.7em {
\lstick{\ket{0}}&\gate{R^X_{\theta_1}}&\ctrl{1}&\meter\\
\lstick{\ket{0}}&\qw&\gate{R^Y_{\theta_2}}&\meter\\
}
&
&\;\;\;\;\;\Qcircuit @C=.7em @R=.7em {
\lstick{\ket{0}}&\gate{R^X_{\theta_1}}&\ctrl{1}&\meter\\
\lstick{\ket{1}}&\gate{R^X_{\theta_1}}&\gate{R^Y_{\theta_2}}&\meter\\\\
(A) \quad\quad\quad\quad\quad\quad\quad\quad\quad\quad (B) \quad\quad\quad\quad\quad
}
\end{align*}
Here, $CR_Y^{0,1}$ is a controlled $Y$ rotation with control qubit $0$ and target qubit $1$, $R_X^q(\theta_1)$ is a rotation of qubit $q$ around the $X$ axis, and the rotation about the $j$ axis is $R_{\sigma_j}(\theta) = e^{-i\theta \sigma_j/2}$ with Pauli matrices $\sigma_j$. 
\newedit{Note that $\theta_3$ is a \textit{fictitious} parameter which corresponds to the global phase. This is present only so that the evolution of the other parameters are not constrained to produce an oscillating global phase in time, as recently studied in }Ref.~\cite{2018arXiv181208767Y}.

\begin{figure}[h!]
    \centering
    \subfigure{
        \includegraphics[width=.8\linewidth,trim=0cm 1cm 0cm 0cm,clip]{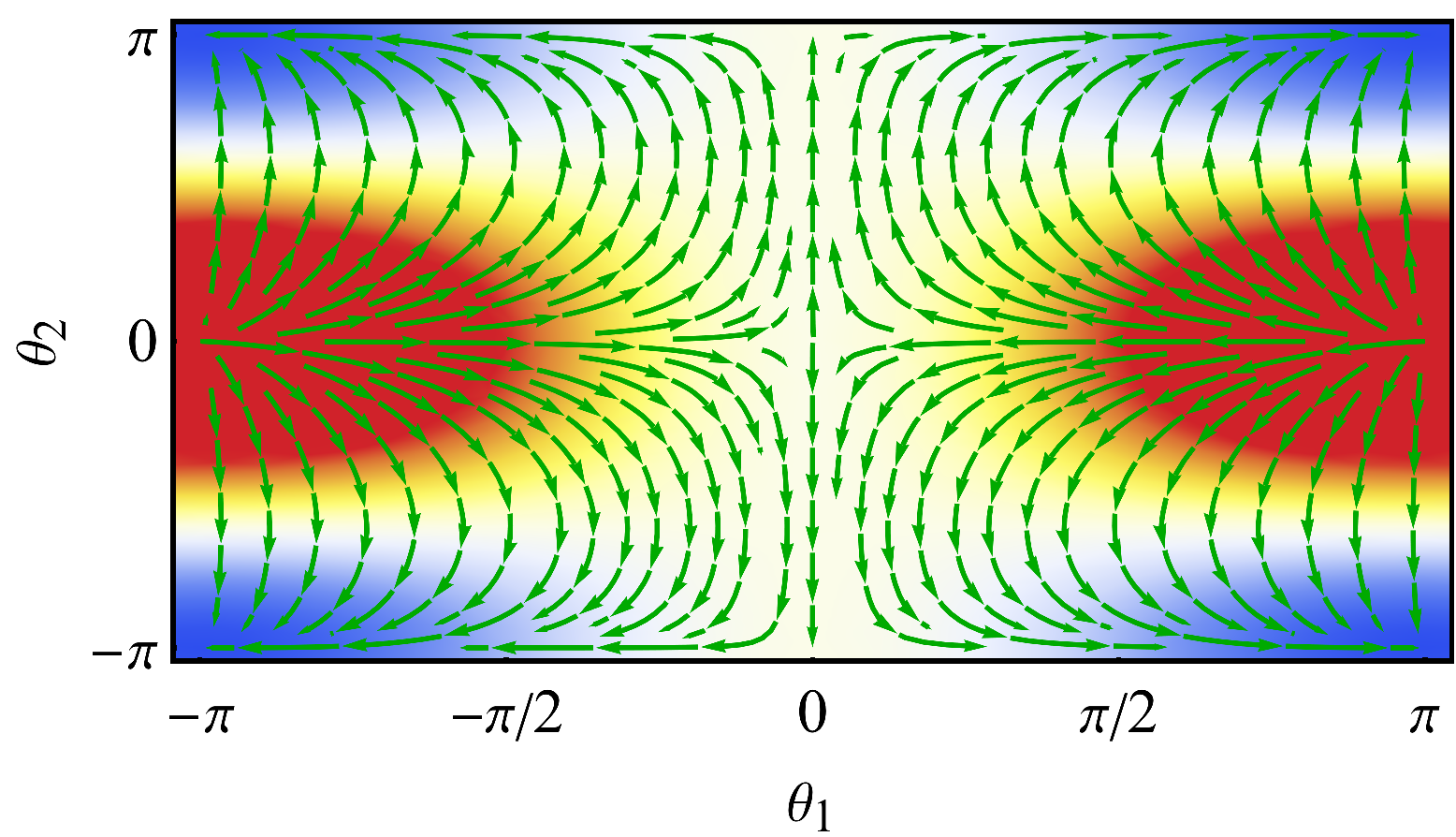}
        \hspace{1cm}
    } \vskip -.2cm
    \setcounter{subfigure}{0}
    \subfigure[ $H_A$, $\ket{\psi_A}$]{
        \includegraphics[width=.8\linewidth,trim=0cm 0cm 0cm .1cm,clip]{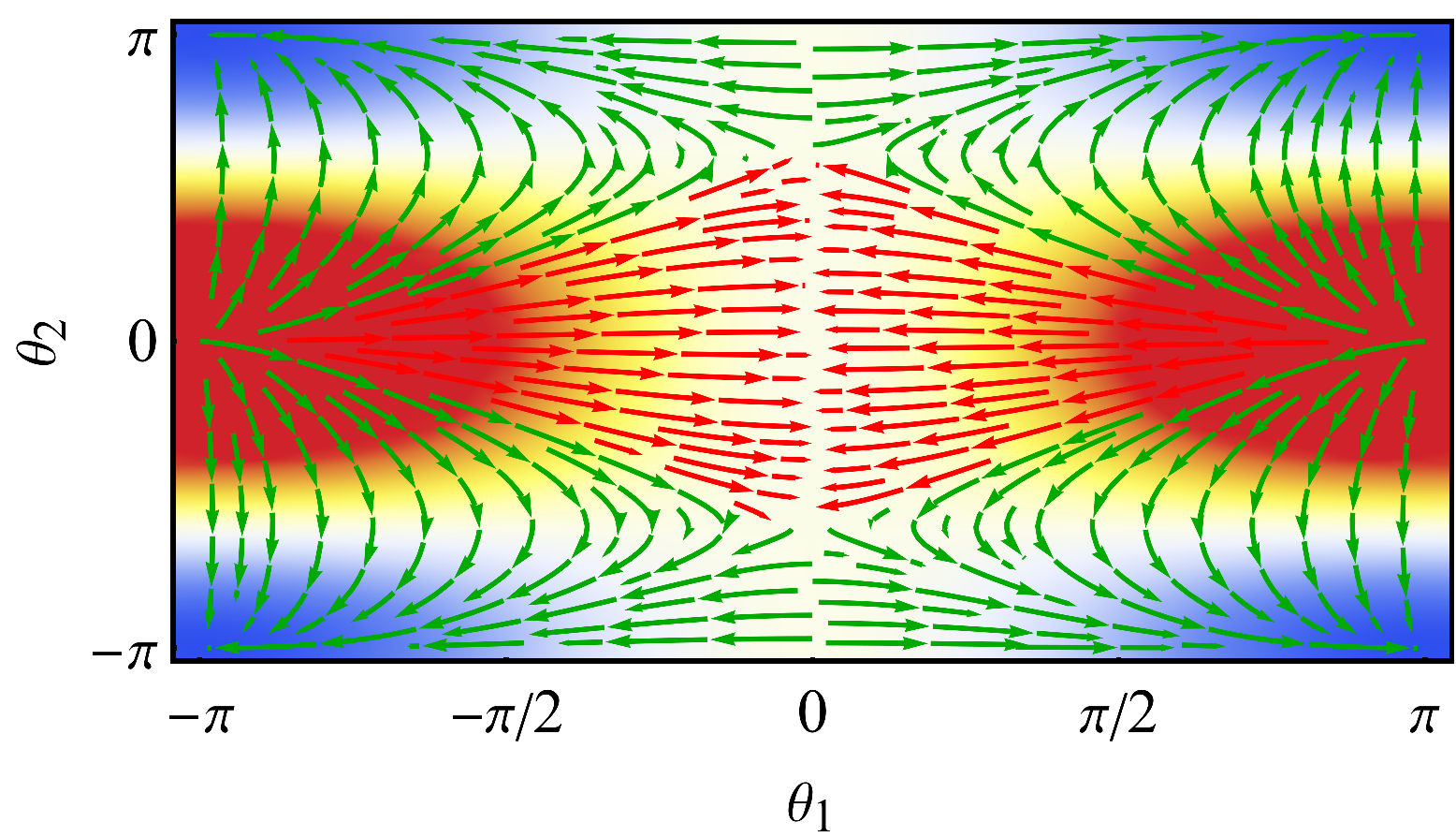}
        \hspace{1cm}
        \label{fig:toy_examples_systemA}
    }
    \subfigure{
        \includegraphics[width=.8\linewidth,trim=0cm 1cm 0cm 0cm,clip]{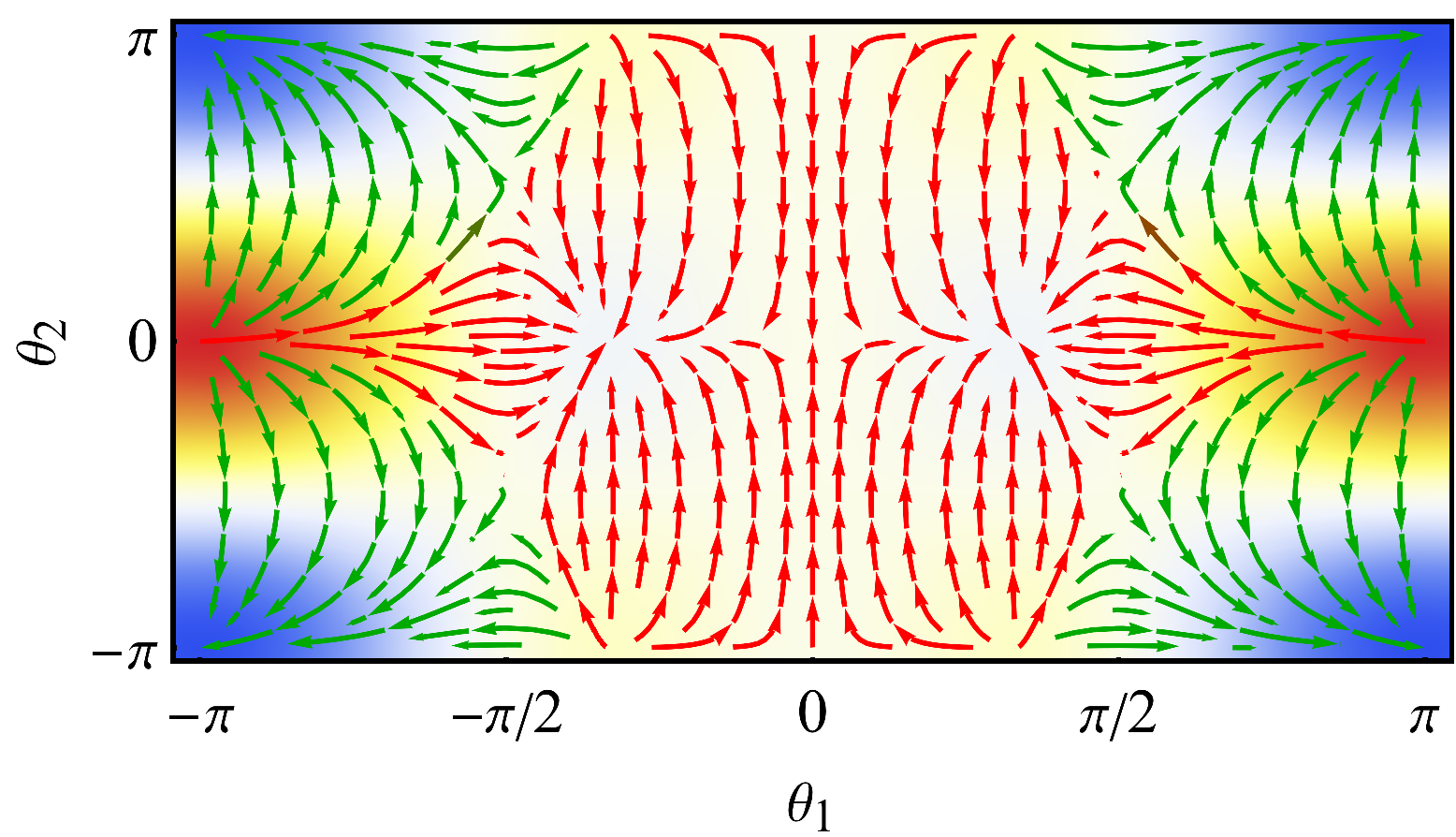}
        \hspace{1cm}
    } \vskip -.2cm
    \setcounter{subfigure}{1}
    \subfigure[ $H_B$, $\ket{\psi_B}$]{
        \includegraphics[width=.8\linewidth,trim=0cm 0cm 0cm .1cm,clip]{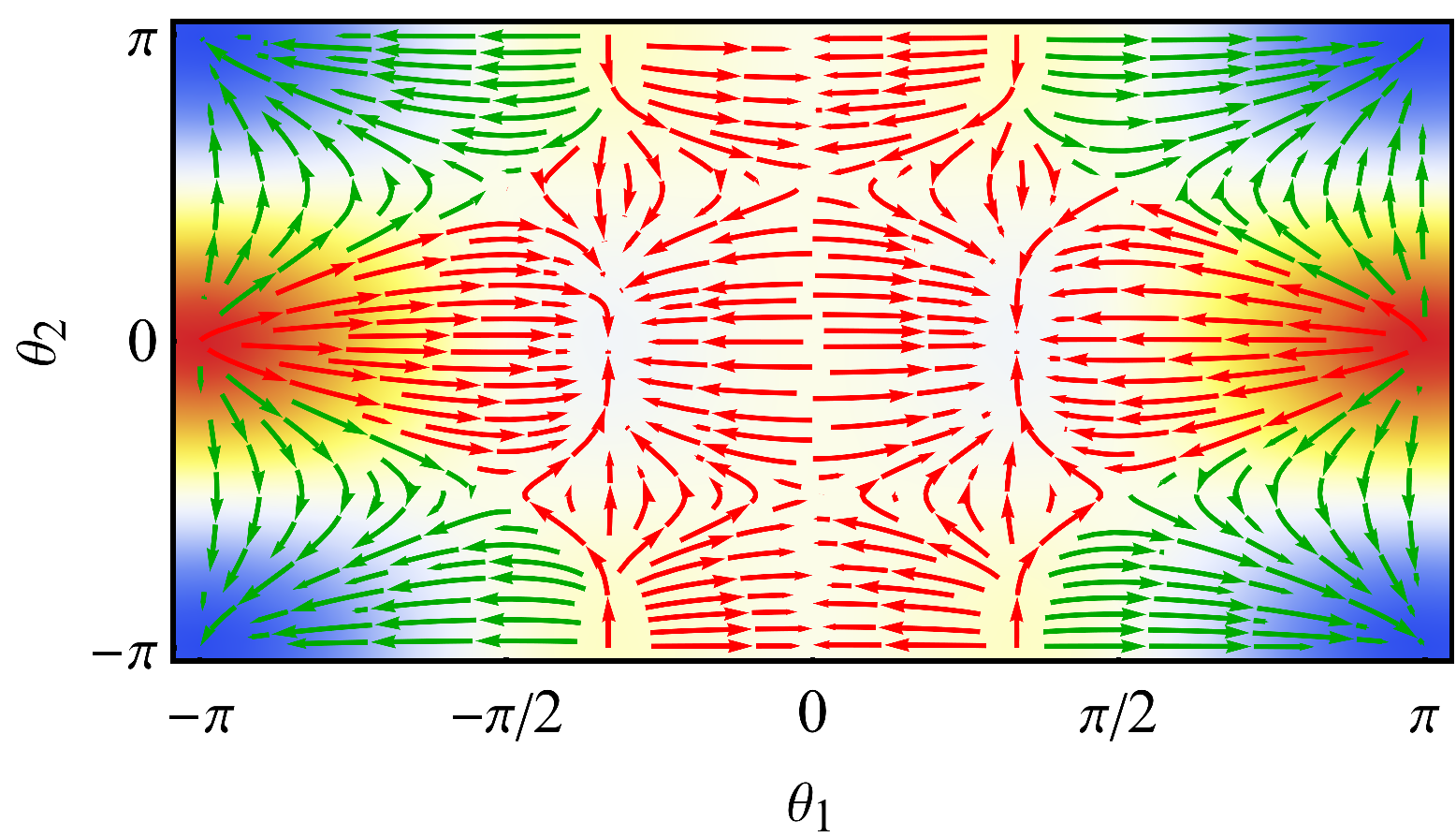}
        \hspace{1cm}
        \label{fig:toy_examples_systemB}
    }
    \caption{
    \newedit{
        Comparison of variational imaginary time (top plot in each panel) and gradient descent (bottom plot in each panel) discovering the ground state in toy systems $A$ (top panel) and $B$ (bottom panel). The background colour indicates the energy $\braket{\psi(\theta_1,\theta_2,\theta_3)|H|\psi(\theta_1,\theta_2,\theta_3)}$ with red and blue corresponding to the global maximum and ground state energies, respectively. The arrows indicate the trajectories of the methods, and are coloured green if they converge to the true ground state, and red otherwise.
        While imaginary time avoids all local minima in system $A$, both methods can become trapped in local minima for the adversarial system $B$.
    }
    }
    \label{fig:toy_examples}
\end{figure}

We study the ability of variational imaginary time and gradient descent to navigate the energy landscapes of the toy systems $A$ and $B$, and present the results in Figure~\ref{fig:toy_examples}. 
Figure~\ref{fig:toy_examples_systemA} \newedit{
shows imaginary time robustly discovering the global minima of the energy landscape of system $A$ , while gradient descent becomes trapped in local minima.
}
Figure~\ref{fig:toy_examples_systemB} \newedit{shows imaginary time performing comparably to gradient descent for system $B$, despite it being only slightly more complicated than system $A$. This shows that it is still possible for imaginary time evolution to become trapped in local minima, when the ansatz is not sufficiently powerful for certain Hamiltonians.\\
}

\begin{figure}[t]\centering
{\includegraphics[width=.45\textwidth]{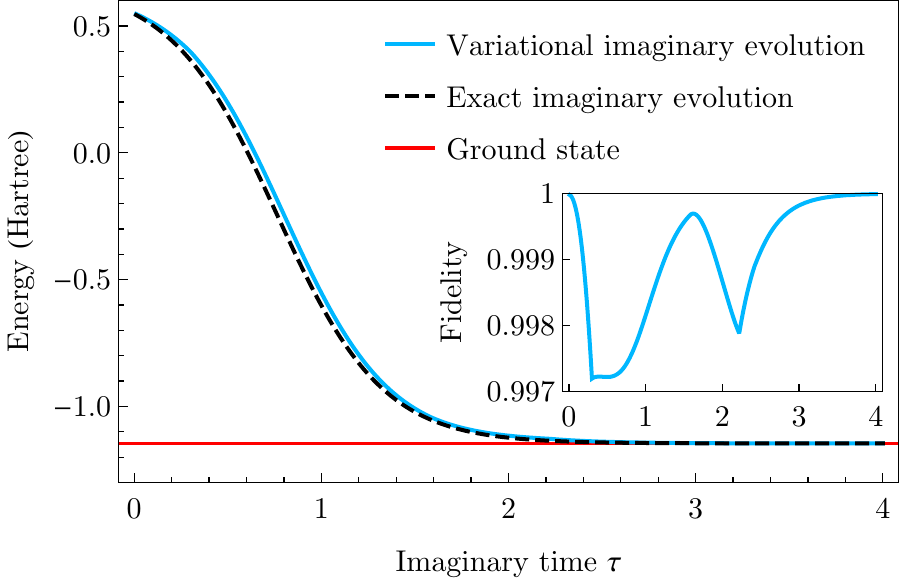}}
 \caption{Simulations of H$_2$ with random initial parameters and timestep $\delta \tau = 0.01$. The red line is the exact ground state energy. The dashed black line is the exact imaginary time evolution. The blue line is the variational imaginary time evolution. {The inset plot shows the fidelity of variational imaginary time to true imaginary time evolution.}
 Here we consider an internuclear distance of $R = 0.75~\mathrm{\AA}$. 
 \newedit{The inset plot and main plot share the same $x$ axis label}. 
} \label{Fig:simulationRes}
\end{figure}

\noindent\textit{Simulation of H$_2$ and LiH.}
We use our method to find the ground state energy of the H$_2$ and LiH molecules in their minimal spin-orbital basis sets. 
We map the molecular fermionic Hamiltonians to qubit Hamiltonians using the procedure described in the Supplementary Materials. The H$_2$ Hamiltonian acts on two qubits, \newedit{and considers the space of two electrons in four spin-orbitals. The LiH Hamiltonian acts on eight qubits, and considers an active space of  two electrons in eight spin-orbitals.} There are numerous possible choices for the ansatz circuit; we use a universal ansatz for H$_2$~\cite{peruzzo2014variational} and an \newedit{ansatz inspired by the low-depth circuit ansatz}~\cite{dallaire2018low} \newedit{for LiH}, as shown in the Supplementary Materials. The simulation results for H$_2$ are shown in Fig.~\ref{Fig:simulationRes}. \newedit{We have used a universal ansatz, which is capable of representing all states along the imaginary time trajectory to confirm that our method can recover true imaginary time evolution, when the ansatz is sufficiently powerful. We attribute deviation from the true evolution to the use of an Euler update rule, and finite step size. Our simulations were able to converge to the ground state in all trials.}

We compare the LiH results to those obtained using the VQE, with gradient descent as the classical optimisation routine. 
We use the low-depth circuit ansatz shown in the Supplementary Materials for our simulation, with 137 parameters. \newedit{This is approximately a quarter the number needed in a universal ansatz.}
\newedit{We  consider starting from a good initial state (the Hartree-Fock state for LiH), and also random initial states. We believe that the latter simulations provide a more thorough test of both methods.}

\begin{figure}[h]
    \centering
    \subfigure{
    \includegraphics[width=0.46\textwidth]{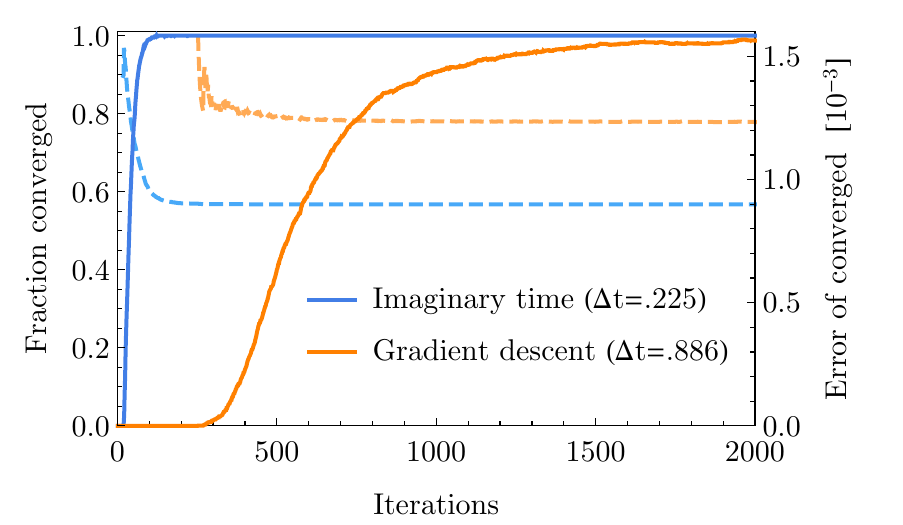}
    } %
    \subfigure{
    \includegraphics[width=0.46\textwidth]{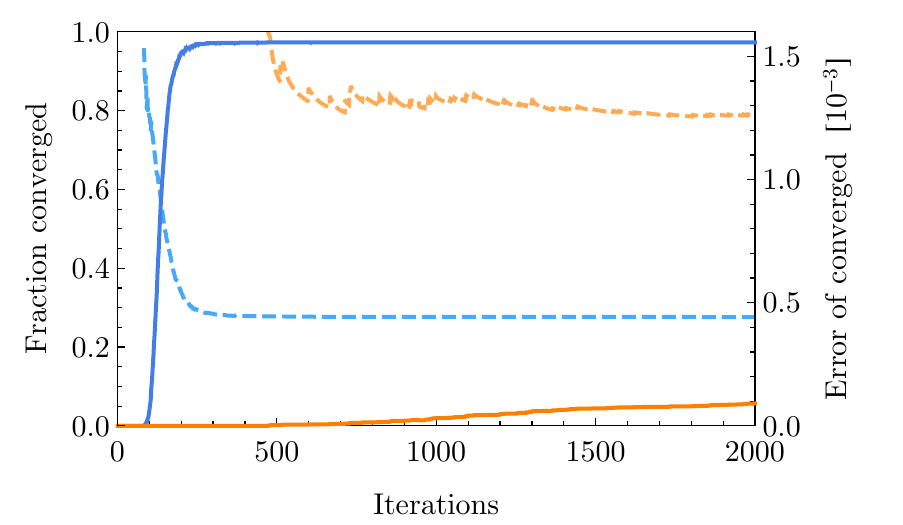}
    }
    \caption{ 
    \newedit{
    Noise-free simulations of LiH at an internuclear distance of $R = 1.45~\mathrm{\AA}$. Simulations in the top plot begin in a small random perturbation (of at most, $\Delta \theta_j = \pi/50$) from the Hartree-Fock state. Simulations in the bottom plot begin with uniformly random parameters. The solid lines (against the left axis) indicate the fraction of $1280$ simulations which, by the given iteration, have converged to within $1\,$mHartree of the true ground state. The dashed lines (against the right axis) indicate the average proximity to the true ground state of only the so-far converged simulations. Imaginary time and gradient descent use their maximum stable timesteps.
    }
    }\label{fig:LiHGroundStates}
\end{figure}

{
We use the maximum stable stepsize $\delta \tau$ for each method such that energy monotonically decreases in the first 200 iterations. \newedit{The stable timestep for imaginary time was 0.225, and for gradient descent it was 0.886}. Fig.~\ref{fig:LiHGroundStates} shows the imaginary time method outperforming gradient descent. It is able to locate the ground state more quickly, and accurately. This advantage is most noticeable for the case of random start states, where the obtained convergence rate is significantly higher than gradient descent. This may be most relevant for solving optimisation problems using the QAOA algorithm, where it is often harder to motivate a good initial starting state.}

\FloatBarrier

\begin{figure}[h]
    \centering
    \includegraphics[width=0.46\textwidth]{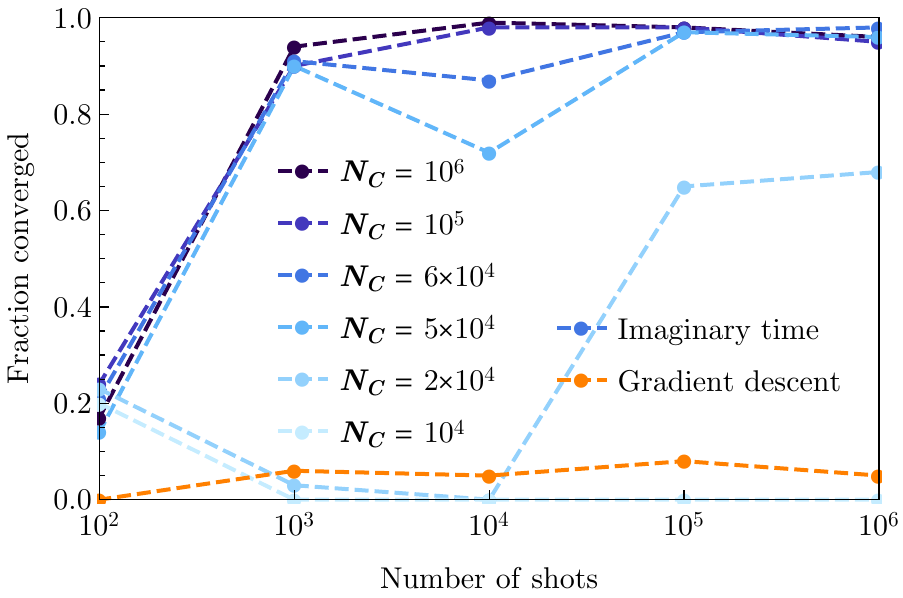}
    \caption{
    \newedit{
    Simulations of LiH in the presence of a $10^{-4}$ error rate per gate and varying amounts of shot noise. Each point indicates the fraction of $100$ trials which, after $2000$ iterations from uniformly random initial parameter states, finished within $1\,$mHartree of the true ground state of LiH. For imaginary time (the blue lines), the horizontal axis indicates the number of shots $N_A$ used in sampling each element of the coefficient matrix $A$, every iteration. The number of shots $N_C$ used in measuring each Hamiltonian \textit{term} in each element of the gradient vector, as employed by imaginary time evolution, is varied between the blue lines. For gradient descent, the horizontal axis is $N_C$.
    }
    }\label{fig:FewerMeasurements}
\end{figure}

It is natural to question whether the resource requirements of variational imaginary time evolution are comparable to those of gradient descent. We assess this by performing a simple resource estimation, and by examining its sensitivity to shot noise and to gate errors within the quantum computer.
At each iteration, populating the gradient vector requires $\mathcal{O}(N_C N_H N_p)$ measurements, where $N_C$ is the number of measurements required to ascertain a Hamiltonian term to the required precision, $N_H$ is the number of terms in the Hamiltonian, and $N_p$ is the number of parameters used in the ansatz. For imaginary time, the total cost is $\mathcal{O}(N_C N_H N_p + N_p^2 N_A)$ where $N_A$ is the number of measurements required to ascertain an element of the $A$ matrix to the required precision. Sensible ans\"atze typically have fewer parameters $N_p$ than there are Hamiltonian terms $N_H$ (in our LiH simulations, $N_p = 137$ and $N_H = 181$). If the number of Hamiltonian terms is considerably larger than the number of parameters used, then the additional cost $\mathcal{O}(N_p^2 N_A)$ of imaginary time can be dominated by the cost of calculating the gradient vector. While this is not true for our LiH simulations, we find that it is possible to further reduce the cost of imaginary time by using fewer measurements for each term in the $A$ matrix than for each element of the gradient vector ($N_A \ll N_C$), while still maintaining imaginary time's superior performance. We demonstrate this in Fig.~\ref{fig:FewerMeasurements}, where we vary the number of measurements used to populate $A$, and simulate the methods under the effect of decoherence. \newedit{The results show that imaginary time can perform significantly better than gradient descent under the presence of noise, even when significantly fewer measurements are made. However, if the gradient is not known to sufficient accuracy ($N_C < 2 \times 10^4$), the reliability of imaginary time evolution cannot be improved by increasing $N_A$, and can even perform less effectively than gradient descent.} Combined with imaginary time's faster convergence and the tendency of gradient based methods to become trapped in local minima, we expect finding the ground state to require substantially fewer measurements using imaginary time than gradient descent.\\

\noindent\textbf{DISCUSSION}\\
In this work, we have proposed a method to efficiently simulate imaginary time evolution using hybrid quantum-classical computing. 
We have applied our method to finding the ground state energy of quantum systems, and have tested its performance on H$_2$ and LiH. 
As imaginary time evolution outperformed gradient descent at this task, we believe our method provides a competitive alternative to conventional classical optimisation routines. We will examine this further in future work.
We expect that our method would also be suitable for solving general optimisation problems, in conjunction with the QAOA, especially given its performance with randomly chosen initial states.

Our method can also be used to prepare a thermal (Gibbs) state, $\rho_T = e^{-H/T}/\tr[e^{-H/T}]$ of Hamiltonian $H$ at temperature $T$. Sampling from a Gibbs distribution is an important aspect of many machine learning algorithms, and so we believe that our method is applicable to problems in quantum machine learning. Moreover, while previous methods to prepare the Gibbs state~\cite{temme2011quantum,PhysRevLett.108.080402} require long gate sequences (and hence, fault tolerance), our method can be implemented using a shallow circuit. Our algorithm can also be combined with recently proposed error mitigation techniques~\cite{Li2017,PhysRevLett.119.180509,endo2017practical,mcardle2018mitigated}, and so is suitable for current quantum hardware.

Although exact imaginary time evolution deterministically propagates a good initial state to the ground state in the limit that $\tau \rightarrow \infty$, our variational method may still converge to higher energy states, if the chosen ansatz is not sufficiently powerful. In future work, we will investigate how our method may be optimally applied to a variety of tasks in chemistry, optimisation and machine learning. This will include developing suitable ans\"atze for a range of problems. \\

\noindent\textbf{DATA AVAILABILITY}\\
The data that support the findings of this study are available from the corresponding author upon reasonable request.\\

\noindent\textbf{CODE AVAILABILITY}\\
The code that supports the findings of this study are available from the corresponding author upon reasonable request.\\

\noindent\textbf{ACKNOWLEDGEMENTS}\\
This work was supported by BP plc and by the EPSRC National Quantum Technology Hub in Networked Quantum Information Technology (EP/M013243/1). YL is supported by National Natural Science Foundation of China (Grant No. 11875050) and NSAF (Grant No. U1730449). SB and YL thank Sergey Bravyi for suggesting an investigation into the imaginary time variant of the algorithm in Ref.~\cite{Li2017}. SE is supported by Japan Student Services Organization (JASSO) Student Exchange Support Program (Graduate Scholarship for Degree Seeking Students).\\

\noindent\textbf{AUTHOR CONTRIBUTIONS}\\
S.M. and X.Y. conceived of the idea, developed the theory, and wrote the manuscript.
T.J., S.M., and X.Y. carried out the numerical simulation.
S.E. and Y.L. contributed to the theory.
S.B. and X.Y. supervised the project.
All authors discussed the results and contributed to the writing of the manuscript.\\
S.M. and T.J. contributed equally to this work.
\\

\noindent\textbf{ADDITIONAL INFORMATION}\\
\noindent\textbf{Competing Interests:}
The authors declare that there are no competing interests.\\

\noindent\textbf{Publisher’s note:} Springer Nature remains neutral with regard to jurisdictional claims
in published maps and institutional affiliations.




\bibliographystyle{apsrev4-1}
\bibliography{ImaginaryTimeBib}

\clearpage
\widetext
\appendix

\section{Supplementary Information}

\section{Variational simulation of imaginary time evolution}
McLachlan's variational principle \cite{McLachlan}, applied to imaginary time evolution, is given by
 \begin{equation}
 	\delta \|({\partial}/{\partial \tau} + H-E_\tau)\ket{\psi(\tau)}\|=0
 \end{equation}
where
\begin{equation}
\|({\partial}/{\partial \tau} + H-E_\tau)\ket{\psi(\tau)}\|=\left(({\partial}/{\partial \tau} + H-E_\tau)\ket{\psi(\tau)}\right)^\dag({\partial}/{\partial \tau} + H-E_\tau)\ket{\psi(\tau)},
\end{equation}
and $E_\tau = \braket{{\psi(\tau)}|H|{\psi(\tau)}}$.
For a general quantum state, McLachlan's variational principle recovers the imaginary time evolution
\begin{equation}
	\frac{\partial \ket{\psi(\tau)}}{\partial \tau} = -(H - E_\tau)\ket{\psi(\tau)}.
\end{equation}

If we consider a subspace of the whole Hilbert space, which can be reached using the ansatz $\ket{\phi(\tau)}=\ket{\phi(\theta_1,\theta_2,\dots,\theta_N)}$, we can project the imaginary time evolution onto the subspace using McLachlan's variational principle. Replacing $\ket{\psi(\tau)}$ with $\ket{\phi(\tau)}$, yields
\begin{equation}
	\begin{aligned}
		\|({\partial}/{\partial \tau} + H-E_\tau)\ket{\phi(\tau)}\|=&\left(({\partial}/{\partial \tau} + H-E_\tau)\ket{\phi(\tau)}\right)^\dag({\partial}/{\partial \tau} + H-E_\tau)\ket{\phi(\tau)},\\
		=&\sum_{i,j}\frac{\partial \bra{\phi(\tau)}}{\partial \theta_i}\frac{\partial \ket{\phi(\tau)}}{\partial \theta_j}\dot{\theta}_i \dot{\theta}_j+ \sum_{i}\frac{\partial \bra{\phi(\tau)}}{\partial \theta_i}(H-E_\tau)\ket{\phi(\tau)}\dot{\theta}_i\\
		+& \sum_{i}\bra{\phi(\tau)}(H-E_\tau)\frac{\partial \ket{\phi(\tau)}}{\partial \theta_i}\dot{\theta}_i +\bra{\phi(\tau)}	(H-E_\tau)^2\ket{\phi(\tau)}. 
\end{aligned}
\end{equation}
Focusing on $\dot{\theta}_i$, we obtain 
\begin{equation}
	\begin{aligned}
		\frac{\partial \|({\partial}/{\partial \tau} + H-E_\tau)\ket{\phi(\tau)}\|}{\partial \dot{\theta}_i}&=\sum_{j}\left(\frac{\partial \bra{\phi(\tau)}}{\partial \theta_i}\frac{\partial \ket{\phi(\tau)}}{\partial \theta_j}+\frac{\partial \bra{\phi(\tau)}}{\partial \theta_j}\frac{\partial \ket{\phi(\tau)}}{\partial \theta_i}\right)\dot{\theta}_j \\
		&+\frac{\partial \bra{\phi(\tau)}}{\partial \theta_i}(H-E_\tau)\ket{\phi(\tau)}+ \bra{\phi(\tau)}(H-E_\tau)\frac{\partial \ket{\phi(\tau)}}{\partial \theta_i}.
\end{aligned}
\end{equation}
Considering the normalisation condition for the trial state $\ket{\phi(\tau)}$, 
\begin{equation}
	\bra{\phi(\tau)}\ket{\phi(\tau)} = 1,
\end{equation}
we have
\begin{equation}
	E_\tau\frac{\partial \braket{\phi(\tau)|\phi(\tau)}}{\partial \theta_i} = E_\tau\left(\frac{\partial \bra{\phi(\tau)}}{\partial \theta_i}\ket{\phi(\tau)}+ \bra{\phi(\tau)}\frac{\partial \ket{\phi(\tau)}}{\partial \theta_i}\right) = 0,
\end{equation}
and the derivative is simplified to 
\begin{equation}
	\begin{aligned}
		\frac{\partial \|({\partial}/{\partial \tau} + H-E_\tau)\ket{\phi(\tau)}\|}{\partial \dot{\theta}_i}&=\sum_{j}A_{ij}\dot{\theta}_j-C_i.
\end{aligned}
\end{equation}
where 
\begin{equation}
\begin{aligned}
	A_{ij} &= \Re\left(\frac{\partial \bra{\phi(\tau)}}{\partial \theta_i}\frac{\partial \ket{\phi(\tau)}}{\partial \theta_j}\right),\\
	C_i &= -\Re\left(\frac{\partial \bra{\phi(\tau)}}{\partial \theta_i} H\ket{\phi(\tau)}\right).
\end{aligned}
\end{equation}

McLachlan's variational principle requires 
\begin{equation}
	\frac{\partial \|({\partial}/{\partial \tau} + H-E_\tau)\ket{\phi(\tau)}\|}{\partial \dot{\theta}_j} = 0,
\end{equation}
which is equivalent to the differential equation of the parameters
\begin{equation}
	\begin{aligned}
		\sum_jA_{ij}\dot{\theta}_j = C_i.
	\end{aligned}
\end{equation}

Denoting $E(\tau) = \bra{\phi(\tau)}H\ket{\phi(\tau)}$, we can show that the average energy always decreases by following our imaginary time evolution algorithm, for a sufficiently small stepsize;
\begin{equation}
\begin{aligned}
	\frac{d E(\tau)}{d \tau} &= \Re\left(\bra{\phi(\tau)}H\frac{d\ket{\phi(\tau)}}{d \tau}\right),\\
	& = \sum_i\Re\left(\bra{\phi(\tau)}H\frac{\partial\ket{\phi(\tau)}}{\partial\theta_i}\dot{\theta}_i\right),\\
	&=-\sum_i C_i \dot{\theta}_i,\\
	& = -\sum_i C_i A^{-1}_{ij}C_j,\\
	&\le0.
\end{aligned}
\end{equation}
The third line follows from the definition of $C_i$; the fourth line follows from the differential equation of $\dot{\theta}$; the last line is true when $A^{-1}$ is positive. 
First, we show matrix $A$ is positive. We consider an arbitrary vector $x = (x_1, x_2,\dots, x_N)^T$, and calculate $x^\dag\cdot A \cdot x$,
\begin{equation}
\begin{aligned}
x^\dag\cdot A \cdot x &= 
	\sum_{i,j}x_i^*A_{ij}x_j,\\
	 &= \sum_{i,j}x_i^*\Re\left(\frac{\partial \bra{\phi(\tau)}}{\partial \theta_i}\frac{\partial \ket{\phi(\tau)}}{\partial \theta_j}\right) x_j,\\
	 &=\sum_{i,j}x_i^*\frac{\partial \bra{\phi(\tau)}}{\partial \theta_i}\frac{\partial \ket{\phi(\tau)}}{\partial \theta_j} x_j + \sum_{i,j}x_i^*\frac{\partial \bra{\phi(\tau)}}{\partial \theta_j}\frac{\partial \ket{\phi(\tau)}}{\partial \theta_i} x_j,\\
\end{aligned}
	\end{equation}
Denote $\ket{\Phi}=\sum_{i}x_i\frac{\partial \ket{\phi(\tau)}}{\partial \theta_i}$, then the first term equals
\begin{equation}
	\sum_{i,j}x_i^*\frac{\partial \bra{\phi(\tau)}}{\partial \theta_i}\frac{\partial \ket{\phi(\tau)}}{\partial \theta_j} x_j = \braket{\Phi|\Phi} \ge 0.
\end{equation}	
Similarly, we can show that the second term is also nonnegative.
Therefore, $x^\dag\cdot A \cdot x\ge0,\forall x$ and $A$ is nonnegative. 
In practice, when $A$ has eigenvalues with value zero, $A$ is not invertible. However, in our simulation, we define the inverse of $A$ to be only the inverse of the nonnegative eigenvalues. 
Suppose $U$ is the transformation that diagonalises $A$, i.e., $G_{i,j} = (UAU^\dag)_{i,j}=0,\forall i\neq j$. Then, we define $G^{-1}$ by
\begin{equation}
	G^{-1}_{i,j} = 
	\left\{
	\begin{array}{ll}
	\frac{1}{G_{i,j}}&	i=j, G_{i,j}\neq0,\\
	0&	i=j, G_{i,j}=0,\\
	0&	i\neq j.
	\end{array}
	\right.
\end{equation}
The inverse of $A$ is thus defined by 
\begin{equation}
	A^{-1} = U^\dag G^{-1} U.
\end{equation}
Because $A$ has nonnegative eigenvalues, $G$, $G^{-1}$, and hence $A^{-1}$ all have nonnegative eigenvalues.

\section{Evaluating $A$ and $C$ with quantum circuits}
\label{sec:evalAC}
In this section, we review the quantum circuit that can efficiently evaluate the coefficients $A$ and $C$ introduced in Ref.~\cite{romero2017strategies,Li2017,dallaire2018low}.

Without loss of generality, we can assume that each unitary gate $U_{i}(\theta_i)$ in our circuit depends only on parameter $\theta_i$ (since multiple parameter gates can be decomposed into this form).
Suppose each $U_i$ is a rotation or a controlled rotation gate, its derivative can be expressed by 
\begin{equation}\label{Eq:derivativeU}
	\frac{\partial U_{i}(\theta_i)}{\partial \theta_i} = \sum_k f_{k,i}U_{i}(\theta_i)\sigma_{k,i},
\end{equation}
with unitary operator $\sigma_{k,i}$ and scalar parameters $f_{k,i}$. The derivative of the trial state is
\begin{equation}
	\frac{\partial \ket{\phi(\tau)}}{\partial \theta_i} = \sum_k f_{k,i} \tilde{V}_{k,i}\ket{\bar{0}},
\end{equation}
with 
\begin{equation}
	\tilde{V}_{k,i} = U_{N}(\theta_N)\dots U_{i+1}(\theta_{i+1})U_{i}(\theta_i)\sigma_{k,i}\dots U_{2}(\theta_2)U_{1}(\theta_1).
\end{equation}
In practice, there are only one or two terms, $f_{k,i} \sigma_{k,i}$, for each derivative.
For example, when $U_i(\theta_i)$ is a single qubit rotation $R_{\theta_i}^Z = e^{-i\theta_i\sigma_Z/2}$, the derivative ${\partial U_{i}(\theta_i)}/{\partial \theta_i} = -i/2\times Ze^{-i\theta_iZ/2}$, and the derivative of the trial state ${\partial \ket{\phi(\tau)}}/{\partial \theta_i}$ can be prepared by adding an extra $Z$ gate with a constant factor $-i/2$. When $U_i(\theta_i)$ is a control rotation such as $\ket{0}\bra{0}\otimes I + \ket{1}\bra{1}\otimes R_{\theta_i}^Z$, the derivative ${\partial U_{i}(\theta_i)}/{\partial \theta_i} = \ket{1}\bra{1}\otimes {\partial R_{\theta_i}^Z}/{\partial \theta_i} =  -i/2\times\ket{1}\bra{1} \otimes Ze^{-i\theta_iZ/2}$. By choosing $\sigma_{1,i} = I\otimes Z$, $\sigma_{2,i} = Z\otimes Z$, $f_{1,i} = -i/4$, and $f_{2,i} = i/4$, we can show Eq.~\eqref{Eq:derivativeU}.

Therefore, the coefficients $A_{ij}$ and $C_i$ are given by
\begin{equation}
	\begin{aligned}
		A_{ij} &= \Re\left(\sum_{k,l} f_{k,i}^*f_{l,j}\bra{\bar{0}} \tilde{V}_{k,i}^\dag \tilde{V}_{l,j}\ket{\bar{0}}\right),\\
		C_i &= \Re\left(\sum_{k,l} f_{k,i}^*\lambda_l\bra{\bar{0}}\tilde{V}_{k,i}^\dag h_l V\ket{\bar{0}}\right).
	\end{aligned}
\end{equation}
All the terms of the summation follow the general form $a{\Re}(e^{i\theta}\bra{\bar{0}}U\ket{\bar{0}})$ and can be evaluated with a quantum circuit.

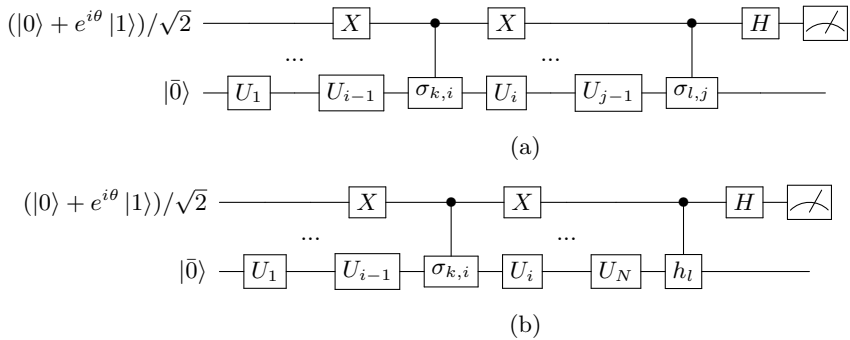
\begin{figure}[hbt]
\begin{align*}
\Qcircuit @C=1em @R=.7em {
\lstick{(\ket{0}+e^{i\theta}\ket{1})/\sqrt{2}}&\qw&\qw&\gate{X}&\ctrl{2}&\gate{X}&\qw&\qw&\ctrl{2}&\gate{H}& \meter\\
&&...&&&&...&\\
\lstick{\ket{\bar{0}}}&\gate{U_1}&\qw&\gate{U_{i-1}}&\gate{\sigma_{k,i}}&\gate{U_{i}}&\qw&\gate{U_{j-1}}&\gate{\sigma_{l,j}}&\qw&\qw\\
}
\end{align*}
(a)
\begin{align*}
\Qcircuit @C=1em @R=.7em {
\lstick{(\ket{0}+e^{i\theta}\ket{1})/\sqrt{2}}&\qw&\qw&\gate{X}&\ctrl{2}&\gate{X}&\qw&\qw&\ctrl{2}&\gate{H}& \meter\\
&&...&&&&...&\\
\lstick{\ket{\bar{0}}}&\gate{U_1}&\qw&\gate{U_{i-1}}&\gate{\sigma_{k,i}}&\gate{U_{i}}&\qw&\gate{U_{N}}&\gate{h_l}&\qw&\qw\\
}
\end{align*}
(b)
\caption{Quantum circuits that evaluate (a) $\Re(e^{i\theta}\bra{\bar{0}}\tilde{V}_{k,i}^\dag \tilde{V}_{l,j}\ket{\bar{0}})$ and (b) $\Re(e^{i\theta}\bra{\bar{0}} \tilde{V}_{k,i}^\dag h_l V\ket{\bar{0}})$. When $\sigma_{k,i}$ is Hermitian, the $X$ gates acting on the ancilla qubit can be also omitted. }\label{Fig:circuitPrac}
\end{figure}

In practice, we do not need to realize the whole controlled-U gate and can instead use a much easier circuit.  For example, for the term $\Re(f_{k,i}^*f_{l,j}\bra{\bar{0}}\tilde{V}_{k,i}^\dag \tilde{V}_{l,j}\ket{\bar{0}})$, we can let $f_{k,i}^*f_{l,j} = ae^{i\theta}$ and 
\begin{equation}
	\bra{\bar{0}} \tilde{V}_{k,i}^\dag \tilde{V}_{l,j}\ket{\bar{0}} = \bra{\bar{0}}U_1^\dag\dots U_{i-1}^\dag \sigma_{k,i}^\dag U_{i}^\dag\dots U_{N}^\dag U_{N}\dots U_{j}\sigma_{l,j}U_{j-1}\dots U_{1}\ket{\bar{0}}.
\end{equation}
Suppose $i < j$, then 
\begin{equation}
	\bra{\bar{0}} \tilde{V}_{k,i}^\dag \tilde{V}_{l,j}\ket{\bar{0}} = \bra{\bar{0}}U_1^\dag\dots U_{i-1}^\dag \sigma_{k,i}^\dag U_{i}^\dag\dots U_{j-1}^\dag \sigma_{l,j}U_{j-1}\dots U_{i} \dots U_{1}\ket{\bar{0}},
\end{equation}
and $\Re(e^{i\theta}\bra{\bar{0}} \tilde{V}_{k,i}^\dag \tilde{V}_{l,j}\ket{\bar{0}})$ can be measured by the circuit in Fig.~\ref{Fig:circuitPrac}. The terms for $C$ can be measured similarly.

\section{Computational chemistry background}
One of the central problems in computational chemistry is finding the ground state energy of molecules. This calculation is classically intractable, due to the exponential growth of Hilbert space with the number of electrons in the molecule. However, it has been shown that quantum computers are able to solve this problem efficiently \cite{aspuru2005simulated}. The Hamiltonian of a molecule consisting of $M$ nuclei (of mass $M_I$, position $\mathbf{R}_I$, and charge $Z_I$) and $N$ electrons (with position $\mathbf{r}_i$) is
\begin{equation}
\begin{aligned}
	H =-\sum_i\frac{\hbar^2}{2m_e}\nabla^2_i  -\sum_I\frac{\hbar^2}{2M_I}\nabla^2_I - \sum_{i,I}\frac{e^2}{4\pi\epsilon_0}\frac{Z_I}{|\mathbf{r}_i-\mathbf{R}_I|}
	+\frac{1}{2}\sum_{i\neq j}\frac{e^2}{4\pi\epsilon_0}\frac{1}{|\mathbf{r}_i-\mathbf{r}_j|}+\frac{1}{2}\sum_{I\neq J}\frac{e^2}{4\pi\epsilon_0}\frac{Z_IZ_J}{|\mathbf{R}_I-\mathbf{R}_J|}.
\end{aligned}
\end{equation}
Because the nuclei are orders of magnitude more massive than the electrons, we apply the Born-Oppenheimer approximation, and treat the nuclei as classical, fixed point charges. After this approximation, the eigenvalue equation we seek to solve (in atomic units) is given by
\begin{equation}\label{ElectronicStructureH}
\begin{aligned}
	\left[-\sum_i\frac{\nabla^2_i}{2} -\sum_{i,I}\frac{Z_I}{|\mathbf{r}_i-\mathbf{R}_I|}+\frac{1}{2}\sum_{i\neq j}\frac{1}{|\mathbf{r}_i-\mathbf{r}_j|}\right]\ket{\psi} = E\ket{\psi},
\end{aligned}
\end{equation}
where $\ket{\psi}$ is an energy eigenstate of the Hamiltonian, with energy eigenvalue $E$. \\

To solve this problem using a quantum computer, we first transform it into the second quantised form. We project the Hamiltonian onto a finite number of basis wave functions, $\{{\phi_p}\}$, which approximate spin-orbitals. Electrons are excited into, or de-excited out of, these spin-orbitals by fermionic creation ($a_p^\dag$) or annihilation ($a_p$) operators, respectively. These operators obey fermionic anti-commutation relations, which enforces the antisymmetry of the wavefunction, a consequence of the Pauli exclusion principle. In the second quantised representation, the electronic Hamiltonian is written as
\begin{equation}\label{Eq:FH}
	H = \sum_{p,q}h_{pq}a^\dag_p a_q + \frac{1}{2}\sum_{p,q,r,s}h_{pqrs}a^\dag_p a^\dag_q a_ra_s,
\end{equation}
with 
\begin{equation}
	\begin{aligned}\label{Integrals}
		h_{pq}&=\int \mathrm{d}\textbf{x} \phi_p^*(\textbf{x}) \left(-\frac{\nabla^2_i}{2} -\sum_{I}\frac{Z_I}{|\mathbf{r}-\mathbf{R}_I|}\right) \phi_q(\mathbf{x}),\\
		 h_{pqrs}&=\int \mathrm{d}\mathbf{x}_1 \mathrm{d}\mathbf{x}_2\frac{\phi_p^*(\mathbf{x}_1) \phi_q^*(\mathbf{x}_2)\phi_s(\mathbf{x}_1)\phi_r(\mathbf{x}_2)}{|\mathbf{r}_1-\mathbf{r}_2|},
	\end{aligned}
\end{equation}
where $\mathbf{x}$ is a spatial and spin coordinate. This Hamiltonian in a molecular orbital basis contains $O(N_{\mathrm{SO}}^4)$ terms, where $N_{\mathrm{SO}}$ is the number of spin-orbitals considered. This fermionic Hamiltonian must then be transformed into a Hamiltonian acting on qubits. This can be achieved using the Jordan-Wigner (JW), or Bravyi-Kitaev (BK) transformations, which are described in Ref.~\cite{BravyiKitaev12}.

\subsection{Hydrogen}
In our simulations, we consider the Hydrogen molecule in the minimal STO-3G basis. This means that only the minimum number of orbitals to describe the electrons are considered. `STO-$n$G' means that a linear combination of $n$~Gaussian functions are used to approximate a Slater-type-orbital, which describes the electron wavefunction. Each Hydrogen atom contributes a single $1S$ orbital. As a result of spin, there are four spin-orbitals in total. We are able to construct the molecular orbitals for H$_2$ by manually (anti)symmetrising the spin-orbitals. These are
\begin{equation}
	\begin{aligned}
		|\phi_0 \rangle = |\sigma_{g \uparrow} \rangle = \frac{1}{\sqrt{\mathcal{N}_G}}(| 1S_{1 \uparrow} \rangle + | 1S_{2 \uparrow} \rangle), \\
		|\phi_1 \rangle = |\sigma_{g \downarrow} \rangle = \frac{1}{\sqrt{\mathcal{N}_G}}(| 1S_{1 \downarrow} \rangle + | 1S_{2 \downarrow} \rangle), \\
		|\phi_2 \rangle = |\sigma_{u \uparrow} \rangle = \frac{1}{\sqrt{\mathcal{N}_U}}(| 1S_{1 \uparrow} \rangle - | 1S_{2 \uparrow} \rangle), \\
		|\phi_3 \rangle = |\sigma_{u \downarrow} \rangle = \frac{1}{\sqrt{\mathcal{N}_U}}(| 1S_{1 \downarrow} \rangle - | 1S_{2 \downarrow} \rangle),
	\end{aligned}
\end{equation}
where the subscripts on the $1S$ orbitals denote the spin of the electron in that orbital, and which of the two hydrogen atoms the orbital is centred on, and $\mathcal{N}_{G/U}$ are normalisation coefficients. By following the procedure in Ref.~\cite{BravyiKitaev12}, the qubit Hamiltonian for H$_2$ in the BK representation can be obtained. This 4 qubit Hamiltonian is given by
\begin{equation}
	\begin{aligned}
	H =&~h_0 I + h_1 Z_0 + h_2 Z_1 + h_3 Z_2 + h_4 Z_0 Z_1 + h_5 Z_0 Z_2 + h_6 Z_1 Z_3 + h_7 X_0 Z_1 X_2 + h_8 Y_0 Z_1 Y_2 \\
		&~+ h_9 Z_0 Z_1 Z_2 + h_{10} Z_0 Z_2 Z_3 + h_{11} Z_1 Z_2 Z_3 + h_{12} X_0 Z_1 X_2 Z_3 + h_{13} Y_0 Z_1 Y_2 Z_3 + h_{14} Z_0 Z_1 Z_2 Z_3 .
	\end{aligned}
\end{equation}
As this Hamiltonian only acts off diagonally on qubits $0$ and $2$ \cite{PRXH2, kandala2017hardware}, it can be reduced to
 \begin{equation}
	\begin{aligned}
	H = g_0I + g_1Z_0 + g_2Z_1 + g_3Z_0Z_1 + g_4Y_0Y_1 + g_5X_0X_1,
	\end{aligned}
\end{equation}
which only acts on two qubits.

\begin{figure}[h]
\centering
\begin{align*}
\Qcircuit @C=1.5em @R=.7em {
\lstick{\ket{0}_1}&\gate{R^Y_{\theta_1}}&\gate{R^Z_{\theta_2}}&\ctrl{1}&\gate{R^Y_{\theta_5}}&\gate{R^Z_{\theta_6}}&\gate{R}&\meter\\
\lstick{\ket{0}_0}&\gate{R^Y_{\theta_3}}&\gate{R^Z_{\theta_4}}&\targ&\gate{R^Y_{\theta_7}}&\gate{R^Z_{\theta_8}}&\gate{R}&\meter\\
}
\end{align*}
\caption{The quantum circuit for preparing the hardware efficient ansatz with eight parameters. Here each $R^{Y}_{\theta}=e^{-i\theta \sigma_Y/2}$, $R^{Z}_{\theta}=e^{-i\theta \sigma_Z/2}$, and the two $R$ gates correspond to the rotation of the measurement.} \label{Fig:H2ansatz}
\end{figure}
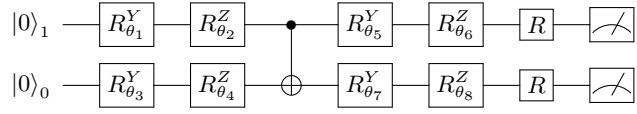

In our work, we consider an internuclear distance of $R = 0.75~\mathrm{\AA}$ and hence $g_0 = 0.2252$, $g_1 = 0.3435$,  $g_2 = -0.4347$ $g_3=0.5716$, $g_4=0.0910$, $g_5=0.0910$. We make use of the universal ansatz~\cite{peruzzo2014variational} shown in Fig.~\ref{Fig:H2ansatz}.

\begin{figure}[h]
\centering
\begin{align*}
\Qcircuit @C=.8em @R=.7em {
\lstick{\ket{0}_2}&\gate{H}&
\gate{X}&\ctrl{1}&
\gate{X}&\qw&\qw&\ctrl{2}&\qw&\qw&\gate{H}&\meter\\
\lstick{\ket{0}_1}&\gate{R^Y_{\theta_1}}&\qw&\gate{Z}&\qw&\gate{R^Z_{\theta_2}}&\ctrl{1}&\qw&\gate{R^Y_{\theta_5}}&\gate{R^Z_{\theta_6}}&\qw&\qw\\
\lstick{\ket{0}_0}&\gate{R^Y_{\theta_3}}&\qw&\qw&\qw&\gate{R^Z_{\theta_4}}&\targ&\gate{Y}&\gate{R^Y_{\theta_7}}&\gate{R^Z_{\theta_8}}&\qw&\qw\gategroup{2}{7}{3}{8}{.7em}{--}
}
\end{align*}
\caption{The circuit to measure $A_{2,7} = \Re\left(\frac{\partial \bra{\phi(\tau)}}{\partial \theta_2}\frac{\partial \ket{\phi(\tau)}}{\partial \theta_7}\right)$. In practice, the gates in the dashed box may be omitted. The other terms of $A$ and $C$ can be measured using similar circuits.
} \label{Fig:H2A27}
\end{figure}
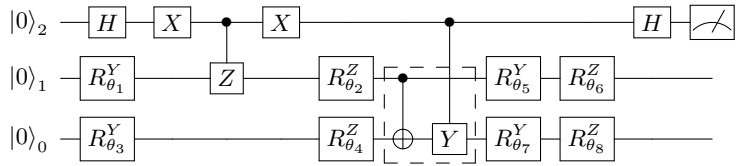

We have eight parameters and an example circuit for measureing $A_{2,7}$ is shown in Fig.~\ref{Fig:H2A27}

\subsection{Lithium Hydride}
In our simulations, we consider Lithium Hydride in the minimal STO-3G basis. The Lithium atom has 3 electrons, and so contributes a $1S, 2S, 2P_x, 2P_y$ and $2P_z$ orbital to the basis, while the Hydrogen atom contributes a single $1S$ orbital. With spin, this makes 12 spin-orbitals in total. However, we are able to reduce the number of spin-orbitals required by considering their expected occupation. This reduces the qubit resources required for our calculation. In computational chemistry, the subset of spin-orbitals included in a calculation is called the active space. 

We first obtain the one electron reduced density matrix (1-RDM) for LiH, using a classically tractable CISD (configuration interaction, single and double excitations) calculation. The 1-RDM for a distance of $1.45~\mathrm{\AA}$ is shown below,
\begin{equation}
	\left(
	\begin{array}{cccccc}
1.99991&	-0.00047&	0.00047&	0&	0&	-0.00120\\
-0.00047&	1.95969&	0.06691&	0&	0&	0.00842	\\
0.00047& 0.06691& 0.00968& 0 & 0& -0.01385\\
0& 0& 0& 0.00172& 0& 0\\
0& 0& 0& 0& 0.00172& 0\\
-0.00120& 0.00842& -0.01385& 0& 0& 0.02728\\
	\end{array}
	\right).
\end{equation}
There are only six rows and columns in the 1-RDM because the spin-up and spin-down orbitals have been combined. The diagonal elements of the 1-RDM are the occupation numbers of the corresponding canonical orbitals (the Hartree-Fock orbitals). In order to reduce our active space, we first perform a unitary rotation of the 1-RDM, such that it becomes a diagonal matrix,
\begin{equation}
	\left(
	\begin{array}{cccccc}
1.99992& 0& 0& 0& 0& 0\\
0& 1.96201& 0& 0& 0& 0\\
0& 0& 0.03459& 0& 0& 0\\
0& 0& 0& 0.00005& 0& 0\\
0& 0& 0& 0& 0.00172& 0\\
0& 0& 0& 0& 0& 0.00172\\
	\end{array}
	\right).
\end{equation}
This gives the 1-RDM in terms of natural molecular orbitals (NMOs). The diagonal entries are called the natural orbital occupation numbers (NOONs). The Hamiltonian of LiH must also be rotated, using the same unitary matrix used to diagonalise the 1-RDM. This is equivalent to performing a change of basis, from the canonical orbital basis to the natural molecular orbital basis.

As can be seen, the first orbital has a NOON close to two, and so is very likely to be doubly occupied. As a result, we `freeze' this core orbital, and consider it to always be doubly filled. We can then remove any terms containing $a_0^\dag, a_0, a_1^\dag, a_1$ from the LiH fermionic Hamiltonian, \newedit{where $0$ and $1$ denote the spin-orbitals that correspond to the core spatial orbital}. We also notice that the fourth spatial orbital has a NOON close to zero. As a result, we assume that this orbital is never occupied by either a spin-up or spin-down electron, and so remove another two fermionic spin-orbital operators from the Hamiltonian. This leaves a fermionic Hamiltonian acting on 8 spin-orbitals. We then map this fermionic Hamiltonian to a qubit Hamiltonian, using the JW transformation. 
All of these steps were carried out using OpenFermion~\cite{mcclean2017openfermion}, an electronic structure package to transform computational chemistry problems into a form that is suitable for investigation using a quantum computer. 

\subsubsection{LiH ansatz}
\begin{figure}[hbt]
\centering
\begin{align*}
\Qcircuit @C=1.em @R=.7em {
\lstick{\ket{1}}&\gate{R^Z}&\gate{R^Y}&\gate{R^X}&\gate{R^Z}&\multigate{1}{U}&\qw&\qw&{\cdots}&&\meter\\
\lstick{\ket{1}}&\gate{R^Z}&\gate{R^Y}&\gate{R^X}&\gate{R^Z}&\ghost{U}&\multigate{1}{U}&\qw&{\cdots}&&\meter\\
\lstick{\ket{0}}&\gate{R^Z}&\gate{R^Y}&\gate{R^X}&\gate{R^Z}&\multigate{1}{U}&\ghost{U}&\qw&{\cdots}&&\meter\\
\lstick{\ket{0}}&\gate{R^Z}&\gate{R^Y}&\gate{R^X}&\gate{R^Z}&\ghost{U}&\multigate{1}{U}&\qw&{\cdots}&&\meter\\
\lstick{\ket{0}}&\gate{R^Z}&\gate{R^Y}&\gate{R^X}&\gate{R^Z}&\multigate{1}{U}&\ghost{U}&\qw&{\cdots}&&\meter\\
\lstick{\ket{0}}&\gate{R^Z}&\gate{R^Y}&\gate{R^X}&\gate{R^Z}&\ghost{U}&\multigate{1}{U}&\qw&{\cdots}&&\meter\\
\lstick{\ket{0}}&\gate{R^Z}&\gate{R^Y}&\gate{R^X}&\gate{R^Z}&\multigate{1}{U}&\ghost{U}&\qw&{\cdots}&&\meter\\
\lstick{\ket{0}}&\gate{R^Z}&\gate{R^Y}&\gate{R^X}&\gate{R^Z}&\ghost{U}&\qw&\qw&{\cdots}&&\meter\\}
\end{align*}
\caption{The general structure of the ansatz used in our LiH simulations. We repeat the circuit structure of the two-qubit rotations block to depth $M=3$. The form of the $U$ gate is $U = e^{i\alpha YX}e^{i\beta XY}e^{i\gamma ZZ}e^{i\delta YY}e^{i\epsilon XX}$. In total there are $(3 \times 5 \times 7) + (4 \times 8) = 137$ parameters.}\label{Fig:generalAnsatz}
\end{figure}
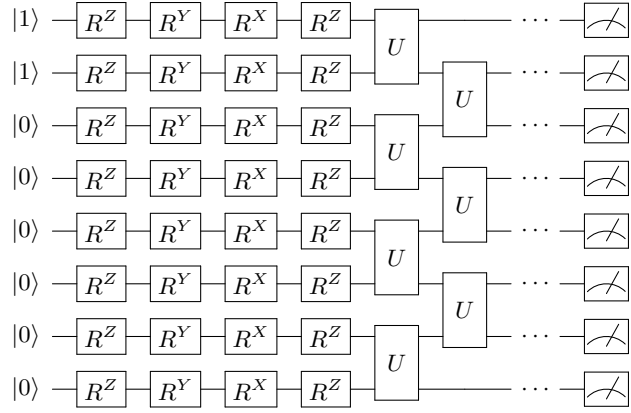

We used an ansatz inspired by the low depth circuit ansatz (LDCA)~\cite{dallaire2018low}. This ansatz is both hardware efficient (in the sense that it only uses nearest neighbour gates), and chemically motivated. The specific ansatz used for simulating LiH is shown in Fig.~\ref{Fig:generalAnsatz}.

\subsubsection{Numerical simulation}

We simulate the aforementioned quantum circuits using the Quantum Exact Simulation Toolkit (QuEST)~\cite{questwhitepaper}.
True ground states are found by diagonalising the considered Hamiltonians with the GNU Scientific Library (GSL)~\cite{GslGnuNumericsLibrary}, which employs a complex form of the symmetric bidiagonalisation and QR reduction method~\cite{gslDiagAlgorithms}.

%
%
%
While our H$_2$ tests simulate the full experimental routine, our LiH tests are optimised by individually computing each ansatz derivative $\partial \ket{\psi(\vec\theta)} / \partial \theta_j$. While obtaining this wavefunction from an experiment would require a number of measurements that grows exponentially with the number of qubits, we can access it directly in our numerical simulations. The observables of the experimental routine are then directly calculated via inner products of these derivative wavefunctions. This allows us to populate the $A$ matrix with only $N_p$ evaluations of the ansatz circuit, in contrast to the ${N_p}^2$ evaluations involved in a full experimental routine. Similarly, the state $\hat{H}\ket{\psi(\theta)}$ can be computed once each iteration, and the $C$ vector populated via the inner product of $\hat{H}\ket{\psi(\theta)}$ with the derivative wavefunctions. This reduces the number of simulated gates by $N_p N_H$ from the full experimental routine described in the main text.

After populating $A$ and $C$, we then update the parameters under the variational imaginary time evolution scheme described in the main text. Since $A \dot{\vec{\theta}} = C \delta t$ is generally underdetermined and leaves us unable to invert $A$, we instead update the parameters under Tikhonov regularisation, which minimises
\begin{equation}
\| C - A \dot{\vec{\theta}} \|^2 + \lambda \| \dot{\vec{\theta}} \|^2.
\end{equation}
Here, the Tikhonov parameter $\lambda$ can be varied to trade accuracy for keeping $\dot{\vec{\theta}}$ small and the parameter evolution smooth. 
We estimate an ideal $\lambda$ at each time-step by selecting the corner of a 3-point L-curve~ \cite{GslGnuNumericsLibrary,gslDiagAlgorithms}, though force $\lambda \in [10^{-4}, 10^{-2}]$.
This is because too large a $\lambda$ over-restricts the change in the parameters and was seen to lead to eventual convergence to non-ground states. Meanwhile, no regularisation ($\lambda=0$) saw residuals in $A^{-1}$ disrupt the monotonic decrease in energy.
Using Tikhonov regularisation affords us a larger time-step than other tested methods, which included LU decomposition, least squares minimisation, singular value decomposition (SVD) and truncated SVD.

We use a basic model of how shot noise and gate errors affect the variational algorithm, by modifying the elements in $A$, and $C$. We here illustrate the procedure for an element of $A$, which is found by $N_{A}$ samples of a binary-outcome measurement with expected value
\begin{align}
a_{ij} = \Re\left( \frac{\partial \bra{\psi(\vec{\theta})}}{\partial \theta_i}  \frac{\partial \ket{\psi(\vec{\theta})}}{\partial \theta_j}  \right) \in \left[ - \frac{1}{4}, \frac{1}{4} \right].
\end{align}
For sufficiently many samples $N_{A}$, the central limit theorem informs $A_{ij}$ is normally distributed.
\begin{align}
A_{ij} \sim N \left( a_{ij}, \frac{1}{N_{A}} \left( 1/16 - a_{ij}^2 \right) \right).
\end{align}
We model the effect of decoherence, viewed as a mixing of each underlying measurement distribution with the fully mixed state, as a skewing of observable expectation values toward their centre (zero). That is, $a_{ij} \to \epsilon a_{ij}$ for some $\epsilon \in [0, 1]$. We assume an experimental error rate of $10^{-4}$ per gate, and approximate the state after $D$ gates to differ from the noise-free state $\rho$ as
\begin{align}
\rho_{\text{final}} = \left(1-10^{-4}\right)^D \rho + \left(1 - \left(1-10^{-4}\right)^D\right) \mathds{1}.
\end{align}
For our LiH simulations $D \approx 100$, so the effect of gate error is to skew our expected values by $\epsilon \approx 0.99$. We follow a similar procedure to introduce noise into the other components of the simulation.

\end{document}